# On the origin of the anomalous compliance of dealloying-derived nanoporous gold

B.-N. D. Ngô [a, *], B. Roschning [b], K. Albe [c], J. Weissmüller [a, b], J. Markmann [a, b]

[a] *Institute of Materials Research, Materials Mechanics, Helmholtz-Zentrum Geesthacht, Geesthacht, Germany*
[b] *Institute of Materials Physics and Technology, Hamburg University of Technology, Hamburg, Germany*
[c] *Division of Materials Modeling, Institute of Materials Science, TU Darmstadt, Darmstadt, Germany*

## Abstract

The origin of the anomalously compliant behavior of nanoporous gold is studied by comparing the elasticity obtained from molecular dynamics (MD) and finite element method (FEM) simulations. Both models yield a compliance, which is much higher than the predictions of the Gibson-Ashby scaling relation for metal foams and thus confirm the influence of other microstructural features besides the porosity. The linear elastic FEM simulation also yields a substantially stiffer response than the MD simulation, which reveals that nonlinear elastic behavior contributes decisively to the anomalous compliance of nanoporous gold at small structure size.

## 1. Introduction

Nanoporous metals, and specifically nanoporous gold (NPG), made by dealloying are emerging as model systems for understanding the mechanical behavior of small-scale solids and of nanomaterials [1–17]. The materials exhibit a uniform microstructure in the form of a network of nanoscale "ligaments", whose size can be adjusted to values between a few and several hundred nanometers [4]. Macroscopic samples typically exhibit highly reproducible mechanical behavior. Their excellent deformability under compression enables experiments that provide access to strain-rate sensitivity [10], elastic and plastic Poisson's ratios [13,16], as well as the evolution of flow stress and stiffness during the plastic densification of the network [12,14,17,18]. Of particular relevance is the observation of strong variations of strength as well as stiffness with ligament size [1,2,6-8,10,11,14]. This finding connects experiments on NPG to topics of interest in the field of small-scale plasticity. Yet, exploiting this connection presupposes that the influence of the network's microstructural features on the mechanical behavior can be disentangled from size effects. This issue is controversially discussed and one aspect of that controversy motivates our work.

As an apparent inconsistency, the early micron-scale tests on NPG indicated a substantially higher strength than the more recent millimeter-scale tests [14]. The tests on macroscopic samples also revealed an anomalously high compliance [14,18]. This was explained in two fundamentally different ways.

Ngô et al. [18] carried out molecular dynamics (MD) simulations, which model the mechanical behavior in quite good agreement with experiments, and proposed a nonlinear elastic response and specifically shear softening in the bulk of the nanoscale ligaments as the origin of the enhanced compliance. Jin et al. [15] and Mameka et al. [14], in contrast, independently emphasized that the anomalous compliance and low strength of NPG are compatible with a conventional, bulk-like elastic response in each ligament, if disorder or defects at the network level are admitted. Indeed, Huber et al. [12] and Roschning et al. [17] have shown that disorder in the array of connecting nodes substantially reduces the stiffness. A less-than-ideal connectivity of the network – for instance in the form of "broken ligaments" – would act similarly [14,15,19,20].

Here, we present a simulation study that is designed to discriminate between these two opposing views. First, we explore the stiffness of a given network structure in a fully atomistic simulation [18] that admits a conceivable nonlinear bulk elastic behavior. Then, we feed the microstructural geometry at various states of plastic strain into finite element method (FEM) simulations for studies of the elastic response. The constitutive law used in the FEM simulations is restricted to linear elasticity, with elastic parameters matching to those of the atomistic simulation. Thus, the FEM simulations explore the impact of the microstructure on the elastic response, but explicitly exclude nonlinear elasticity. If one adopts the view of the "network geometry" put forward in Refs. [12,14,15], both simulations should yield an identical result for the effective macroscopic stiffness. By contrast, the "nonlinear bulk elasticity" picture of Ref. [18] implies that the atomistic simulation gives a larger anomalous compliance. Thus, this comparison provides a basis for assessing the two distinct views.

\* Corresponding author.
Email address: dinh.ngo@hzg.de (B.-N. D. Ngô)

## 2. Methods

### 2.1. Microstructure generation

A nanoporous microstructure was created by mimicking spinodal decomposition of a binary solid solution on an FCC lattice with 144 × 144 × 144 lattice spacings (587.52 × 587.52 × 587.52 Å) and <100> edges, using the Metropolis Monte Carlo algorithm with an Ising-type Hamiltonian and periodic boundary conditions. After the phase separation, the atoms of one component were removed, leaving behind an interconnected network of atoms with the solid fraction $\varphi = 0.302$. The alpha-shape surface reconstruction algorithm [21,22] gives the surface-to-volume ratio $\alpha = 0.877$/nm. Using the conversion rule given in [18] (cf. Equation (1) therein), a ligament size, $L$, of 3.76 nm is obtained for the as-prepared NPG structure.

### 2.2. Molecular dynamics simulation

MD simulations using an EAM potential for gold [23] were carried out with the open-source simulation code LAMMPS [24]. For the visualization and analysis we used the open-source software OVITO [25]. The simulation procedure started with a relaxation of atom positions via an energy minimization using the conjugate gradient algorithm. At convergence, the relative change in energy and the specified force tolerance were less than $10^{-12}$ and $10^{-4}$ eV/Å, respectively. After this initial, athermal relaxation, the NPG sample was thermally relaxed for 1 ns at 300 K with no load applied.

Uniaxial compression tests were simulated with a strain rate of $10^8$/s at 300 K. Load/unload sequences were interspersed. The stresses normal to the strain axis were kept at 0 bar. In all simulations, temperature and pressure were controlled by a Nosé–Hoover thermostat and barostat [26–29]. For more details see the report of our earlier, analogous work in Ref. [18].

The EAM interatomic potential exhibits the following linear elastic parameters at small strains and zero temperature: $C_{11} = 183$ GPa, $C_{12} = 159$ GPa, $C_{44} = 45$ GPa [23]. The linear elastic constitutive law of the FEM simulation used identical values.

### 2.3. Finite element simulation

To transfer the microstructures in MD to the FEM study, the surface at each strain state was reconstructed by the alpha-shape method [22] via the Edelsbrunner's algorithm [21]. This algorithm is based on a Delaunay triangulation of the point cloud representing the atoms [21]. As the surface is unrealistically rough, smoothing was applied to obtain a realistic representation of the mesoscale geometry of the ligaments. The surface models were exported to the commercial FEM software ABAQUS and meshed with $\sim 3.5 \times 10^6$ 10-node quadratic tetrahedron (C3D10) elements [30]. With that number of elements the computation is expensive, but remains manageable, since the restriction to purely elastic behavior drastically reduces the number of iterations. The Delaunay triangulation interpolates the surface at the atomic center points, which are systematically inside any sensible location of the surface in a continuum picture. The accumulated volume of the individual elements of the FEM models corresponded to solid fractions which are less than in the original MD structure; the reduction is by $\Delta\varphi = 3.39 \pm 0.16$ percentage points.

Periodic boundary conditions were applied via linear constraint equations, coupling the nodal displacements on opposing faces of the representative volume element (RVE) in all spatial directions. In order to allow for lateral extension, inhomogeneous constraints were prescribed via dummy nodes which are not attached to any part of the model. Since the input mesh did not provide congruent nodes on the opposing faces of the RVE, the nodes on one of two associated faces were duplicated. Nodal based surfaces of the duplicated node set and the opposing node set were defined and coupled via tie constraints. Linear constraint equations were then defined between the initial node set and the congruent duplicated node set.

An orthotropic linear elastic material model was used with the stiffness parameters derived from the EAM potential (see above). The effective stiffness of the RVE was determined from the net compressive stress (reaction force in load direction divided by current cross section) at the macroscopic strain of 0.02.

### 2.4. Verifying the reconstruction approach

As a verification of our FEM study of the nanoporous structure we have studied the elastic response of network structures with diamond geometry. We created diamond lattices with nominally identical geometries *i.*) from the reconstruction of the atomistic model, using the analogous procedure as for the nanoporous structure, and *ii.*) by a classic design of the network microstructure as in a conventional FEM study. The rationale of this approach is to investigate possible artifacts occurring from surface meshing and smoothing of the atomistic input structure. Since the volume fraction of the atomistic input structure and the reconstructed FEM model are slightly different, we were also interested in how this small difference affects the calculated moduli.

The microstructure here consisted of cylindrical struts, linking spherical nodes on diamond lattice positions. As in a previous study by Huber et al. [12], acute connecting angles at ligament junctions were avoided by setting the node radius to $\sqrt{3/2}$ times the ligament radius.

For the atomistic model, the RVE comprised four diamond lattice unit cells in each spatial direction. The RVE was then inscribed on an FCC lattice, with 144 lattice spacings along each dimension. This resulted in a simulation box edge-length of 58.752 nm and a ligament radius of 2.0 nm. The solid fraction of this atomistic model was $\varphi = 0.302$. Thermal relaxation at 300 K slightly increased the solid fraction to $\varphi = 0.304$. Reconstructing an FEM model by the same procedure as for the NPG structure of Section 2.3 yields a diamond structure with $\varphi = 0.268$ and $Y^{\text{eff}} = 3.289$ GPa.

For the stick-and-ball diamond geometry, a reference diamond structure was created in ABAQUS CAE and meshed with C3D10 elements. The RVE here consists of one unit cell. Congruent nodes on opposing faces allow the direct application of periodic boundary conditions via linear constraint equations. For $\varphi = 0.268$, which is the solid fraction of the FEM diamond structure which we reconstructed based on the atomistic model (see above), we obtain $Y^{\text{eff}} = 3.107$ GPa. This differs only by 6% from the value obtained before of the reconstructed structure.

This comparison implies that our strategy for transferring the microstructures from the atomic-scale data set of the MD simulation to the FEM simulation is appropriate and provides results that agree with more conventional FEM models to within a few percent.

We have also studied the reference diamond structure with $\varphi = 0.308$, which corresponds to the original atomistic disordered structure. Here, the FEM calculation yields $Y^{\text{eff}} = 4.015$ GPa. Thus, the comparison of MD and FEM simulations is primarily affected by the reconstruction and smoothing, since this modifies the volume fraction. The modified $\varphi$ entails a difference of about 23% in $Y^{\text{eff}}$. This may be taken as a rough estimate of the impact of the different solid fractions for the comparison of the MD and FEM results of the nanoporous structure in our study.

## 3. Results

Fig. 1a depicts the microstructure of our NPG sample after the thermal relaxation and prior to the onset of straining. The solid fraction of this relaxed structure is $\varphi = 0.308$, slightly more than the initial value due to the densification during the thermal relaxation [18,31]. The maximum plastic compressive true strain, $\varepsilon = 0.335$, of the MD study increased the solid fraction to $\varphi = 0.415$. A snapshot of that densified microstructure is shown in Fig. 1b.

Fig. 2 shows the compressive stress-strain curve from the MD simulation. The result agrees well with that of Ref. [18], which used identical procedures but a slightly smaller ligament size. Since that reference also found agreement between the MD simulation and experiment, the MD results of the present work may be considered relevant for understanding the experimental observations.

The effective Young's modulus, $Y^{\mathrm{eff}}$, at different strain states was determined as the secant modulus [32] during the unload segments of the stress-strain graph. The evolution of $Y^{\mathrm{eff}}$ during the deformation event of Fig. 2 is shown as the upward-pointing triangles in Fig. 3.

In the MD simulation, $Y^{\mathrm{eff}}$ of the NPG sample starts out at 380 MPa. Several studies (see, e.g., Ref. [8] and the references therein) refer the elasticity of NPG to the Gibson-Ashby scaling relation for the stiffness of foams [33], $Y^{\mathrm{eff}} = Y^{\mathrm{bulk}}\varphi^2$. This relation here predicts $Y^{\mathrm{eff}} \approx 7.4$ GPa for the present sample (this estimate is based on the elastic modulus, $Y^{\mathrm{bulk}} = 78$ GPa, of the nontextured polycrystalline Au calculated by using the Kröner's formulation [34] with the linear elastic constants of the EAM potential given above). Thus, the MD simulation finds the initial Young's modulus about a factor of 19 less than predicted by the scaling relation. This agrees with earlier observations from simulation and experiment [14,18].

The effective Young's modulus increases during the plastic compression. The increase is partly expected because of the densification (i.e. increase in the mean ligament thickness) that accompanies the deformation. For example, the MD data in Fig. 3 are from an interval of compression in which the density increases by a factor of 1.35, from $\varphi = 0.308$ to $\varphi = 0.415$, while the stiffness increases more than 5-fold, from $Y^{\mathrm{eff}} = 380$ MPa to $Y^{\mathrm{eff}} = 1.96$ GPa. Other microstructural changes, for example changes in the network topology [14,19,20,35], might also contribute to the increase of the elastic modulus.

Next, we inspect the effective Young's modulus from the linear-elastic FEM simulation, using the MD microstructures at different plastic deformation states as the initial, stress-free geometry. Snapshots of the FEM meshes generated from the undeformed and the most deformed atomistic NPG sample are shown in Figs. 1c–d.

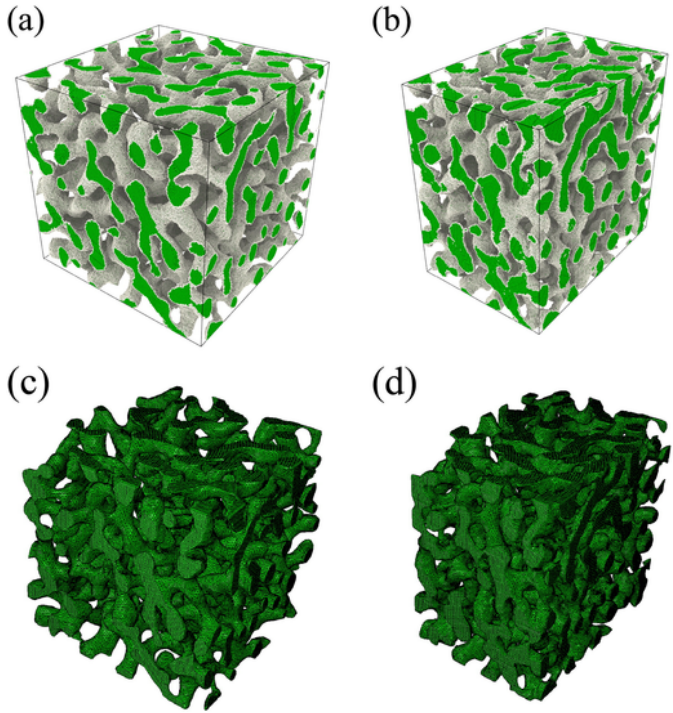

Fig. 1. Geometry of the porous structure. For the molecular dynamics simulation, (a) and (b) show the initial geometry before onset of plastic deformation and the final geometry after deformation to plastic compressive true strain of 0.335, respectively. For the sake of clarity, surface atoms and bulk atoms are coded in different colors. Figures in (c) and (d) depict the finite element meshes reconstructed from the atom coordinates of the structure in (a) and (b), respectively. (For interpretation of the references to color in this figure legend, the reader is referred to the web version of this article.)

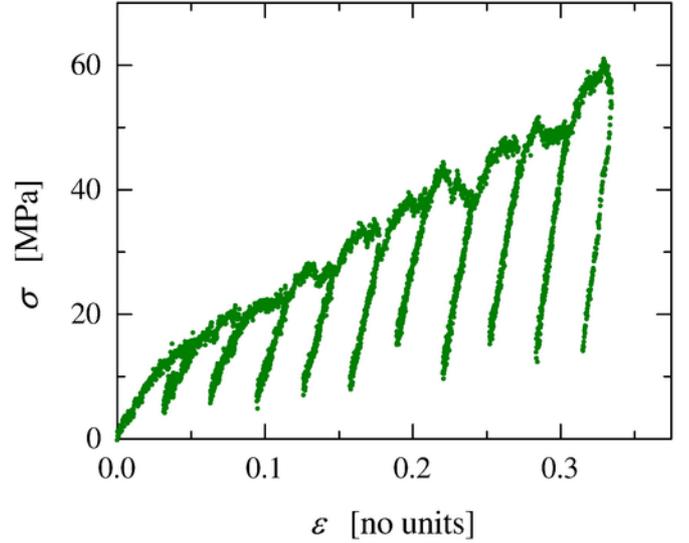

Fig. 2. Mechanical behavior of nanoporous gold during compression by molecular dynamics simulation. Graph of compressive virial stress, $\sigma$, versus compressive true strain, $\varepsilon$. Load unload segments served to monitor the evolution of the effective Young's modulus during plastic densification, see main text.

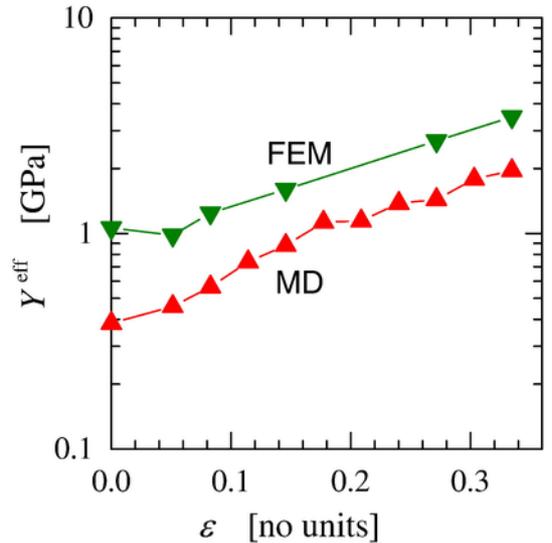

Fig. 3. Evolution of the stiffness during deformation. Effective Young's modulus, $Y^{\mathrm{eff}}$, versus plastic strain, $\varepsilon$. Data from molecular dynamics (MD) and linear-elastic finite element method (FEM) simulations for NPG, as indicated by labels. Note the substantially stiffer behavior of the linear elastic solid, in spite of identical network geometry.

Fig. 3 shows the FEM results for $Y^{\text{eff}}$ alongside the data from MD simulation. We find the FEM value $Y^{\text{eff}} = 1.06$ GPa for the initial configuration. This is again much less than what would be predicted by the Gibson-Ashby scaling law. More importantly, the value of $Y^{\text{eff}}$ of the FEM simulation exceeds that of the MD study by a factor of 2.8. The relative difference remains high throughout the entire set of configurations under investigation.

This difference is highly significant: In view of the practically identical microstructures in the two simulation approaches, the enhanced $Y^{\text{eff}}$ in the linear elastic continuum model that underlies the FEM simulation can only be understood as the consequence of a softer elastic response at the nano- or atomic-scale in the MD simulation. The marginally lesser density of the FEM sample (due to meshing, see above) does not affect this conclusion. In fact, if the density in FEM is to be corrected to match the greater value in MD, the modulus in FEM will increase, implying an even more significant discrepancy.

## 4. Discussion and conclusions

The key finding of our study is that the same network geometry yields relatively higher stiffness when linear elastic behavior is assumed in the FEM and (substantially) lower stiffness when the full nonlinear elastic behavior of the more realistic (EAM-) interatomic potential of the MD is admitted. We also find that both models yield stiffnesses considerably below the prediction of the Gibson-Ashby scaling law that has been found applicable in part of the experimental studies.

We focus first on the last observation, which is also compatible with the low stiffness reported by some experimental studies. Consistent with other simulation studies of NPG, our microstructure was generated by mimicking spinodal decomposition. The geometry appears similar to experimental NPG, yet the notion of similarity is hard to quantify as there is no established link between the mechanical behavior and quantifiable measures for the microstructural topology [14]. It is conceivable that apparently similar microstructures – such as the various spinodal structures in the simulations, the periodic structures that underlie the Gibson-Ashby scaling equation, and experimental NPG made with different dealloying or coarsening parameters – in fact exhibit substantial differences in topology that strongly impact their mechanical behavior [20]. This notion is consistent with our results.

Let us now return to deviations of the local elastic response at the ligament level from the bulk-like linear elastic behavior. The obvious conclusion from our study is that the suggestion of an enhanced compliance at that level is confirmed for the small-scale (∼ 4 nm) structures of the MD study. Ngô et al. [18] motivated this suggestion by the deviatoric component of the surface-induced stresses in high aspect ratio nanostructures and by the classic instability of crystals at high shear stress. The contribution of bulk nonlinear elastic response to the stiffness of nanowires has indeed been confirmed by experiment [36] and density functional theory calculation [37].

It must be emphasized that the nonlinear elastic response is expected only at very small size where the surface-induced stress and the ensuing shear strain are high. The prohibitive computational effort that would be involved in atomistic simulations of realistic volume elements of nanoporous gold with substantially larger ligament size prevents us from studying the size dependence of the elastic behavior by MD. Yet, it appears likely that the high compliance in experimental studies with ligament sizes ≳ 20 nm results exclusively from network geometry effects, for instance differences in the scaled connectivity and/or scaled genus density as suggested in Refs. [14,20] and not from bulk nonlinear behavior.

## Acknowledgments

The authors gratefully acknowledge the computing time granted by the John von Neumann Institute for Computing (NIC) and provided on the supercomputer JURECA [38] at Jülich Supercomputing Centre (JSC), project HHH34. We acknowledge support by German Research Foundation (DFG) via SFB 986 "Tailor-Made Multiscale Materials Systems - $M^3$", projects B2 and B8.